\documentclass[lettersize,journal]{IEEEtran}
\usepackage{graphicx}
\usepackage{booktabs} 
\usepackage{adjustbox}
\usepackage{subcaption}
\usepackage{multirow}
\usepackage{amsmath}
\usepackage{amssymb}
\usepackage{xcolor}
\usepackage{breqn,xspace}
\usepackage{bm}
\usepackage{lipsum,multicol}
\usepackage{makecell}
\usepackage{booktabs}
\usepackage{algorithm}
\usepackage{algorithmic}
\usepackage{amsmath}
\newcommand{\name}{SATSense\xspace}

\ifodd 1
\definecolor{darkgreen}{RGB}{0,200,0}
\newcommand{\rev}[1]{{\color{blue}#1}} 
\else
\newcommand{\rev}[1]{#1}
\fi
\begin{document}
%
\title{\name: Multi-Satellite Collaborative Framework for Spectrum Sensing}
%
%
%
%

\author{Haoxuan~Yuan, Zhe~Chen,~\IEEEmembership{Member,~IEEE,} Zheng~Lin, Jinbo~Peng, Zihan~Fang, Yuhang~Zhong, Zihang~Song,~\IEEEmembership{Member,~IEEE} and~Yue~Gao,~\IEEEmembership{Fellow,~IEEE} 
\IEEEcompsocitemizethanks{\IEEEcompsocthanksitem
H. Yuan, Z. Chen, Z. Lin, J. Peng, Z. Fang, Y. Zhong, X. Wang and Y. Gao are with the School of Computer Science, Fudan University, Fudan University, Shanghai 200438, China (email:hxyuan22@m.fudan.edu.cn; zhechen@fudan.edu.cn; zlin20@fudan.edu.cn; jbpeng22@m.fudan.edu.cn; zhfang19@fudan.edu.cn; zhongyuhang@m.fudan.edu.cn; wangxiong@fudan.edu.cn; gao$\_$yue@fudan.edu.cn ).
\IEEEcompsocthanksitem
Z. Song is with the Department of Engineering, King's College London, Strand, London, WC2R 2LS, United Kingdom (email: zihang.song@kcl.ac.uk). 
}

\thanks{}}

%
%

\markboth{Journal of \LaTeX\ Class Files,~Vol.~14, No.~8, August~2015}%
{Shell \MakeLowercase{\textit{et al.}}: Bare Advanced Demo of IEEEtran.cls for IEEE Computer Society Journals}

\IEEEtitleabstractindextext{%
\begin{abstract}
Low Earth Orbit satellite Internet has recently been deployed, providing worldwide service with non-terrestrial networks. With the large-scale deployment of both non-terrestrial and terrestrial networks, limited spectrum resources will not be allocated enough. Consequently , dynamic spectrum sharing is crucial for their coexistence in the same spectrum, where accurate spectrum sensing is essential. However, spectrum sensing in space is more challenging than in terrestrial networks due to variable channel conditions, making single-satellite sensing unstable. Therefore, we first attempt to design a collaborative sensing scheme utilizing diverse data from multiple satellites. However, it is non-trivial to achieve this collaboration due to heterogeneous channel quality, considerable raw sampling data, and packet loss. To address the above challenges, we first establish connections between the satellites by modeling their sensing data as a graph and devising a graph neural network-based algorithm to achieve effective spectrum sensing. Meanwhile, we establish a joint sub-Nyquist sampling and autoencoder data compression framework to reduce the amount of transmitted sensing data. Finally, we propose a contrastive learning-based mechanism compensates for missing packets. Extensive experiments demonstrate that our proposed strategy can achieve efficient spectrum sensing performance and outperform the conventional deep learning algorithm in spectrum sensing accuracy.

\end{abstract}

}

\maketitle

\IEEEdisplaynontitleabstractindextext

%
\IEEEpeerreviewmaketitle


\section{Introduction}\label{sec:intro}

Low Earth Orbit (LEO) satellites travelling at low orbit altitudes facilitate low end-to-end latency when providing Internet service to ground users.
With the rapid development of satellite manufacturing and launch technologies, large-scale LEO satellite constellations have been constructing~\cite{lin2021path,xiao2022leo}. For example, SpaceX's Starlink and Amazon's Kuiper have been under construction, which aim to offer high-speed and low-latency Internet service by deploying tens of thousands of LEO satellites ~\cite{okati2020downlink,mcdowell2020low,pachler2021updated,lin2023fedsn}.
\begin{figure}
    \centering
    \includegraphics[width=1\linewidth]{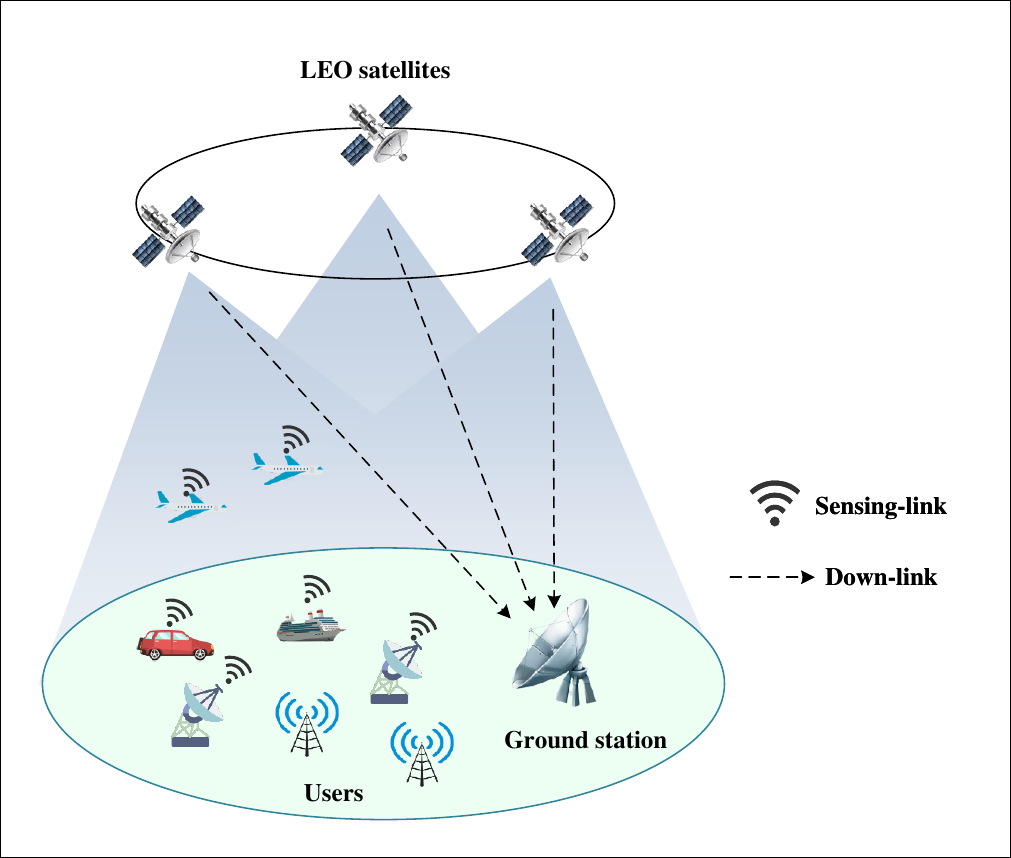}
    \caption{A scenario of collaborative spectrum sensing with multiple LEO satellites.}
    \label{fig:teaser}
\end{figure}
However, large-scale deployment of non-terrestrial networks intensify the spectrum competition with terrestrial networks. Specifically, there may exist overlapped spectrum occupied by LEO satellites and terrestrial networks owned by different organizations, resulting in spectrum interference between wireless communication devices on the ground and LEO satellites~\cite{spacexKuiper2021, an2016outage, an2018performance, nokia,lin2021spatial}. 
Given the limited spectrum resources, achieving efficient spectrum sharing between LEO satellites and terrestrial communication systems is of critical importance. 


Dynamic spectrum sharing facilitates the coexistence of different types or standards of networks by permitting secondary users (SUs) to utilize unused spectrum resources that are not currently occupied by primary users (PUs)~\cite{zhao2007survey}. As a key to spectrum sharing, spectrum sensing technology enhances the utilization efficiency of spectrum resources by monitoring the wireless spectrum to identify unused spectrum resources. Currently, there are many studies related to ground sensing nodes~\cite{liu2019deep, uvaydov2021deepsense,baldesi2022charm,zhang2022machine}. Compared to ground-based sensing nodes, LEO satellites offer advantages as sensing nodes due to their broad coverage and elevated vantage points for observation. 

Nonetheless, the quick movement of satellites and the long distance of the satellite-to-ground link make the communication channel highly susceptible to fluctuating environmental factors, such as electromagnetic disturbances and atmospheric decay, resulting in inconsistent and ever-changing channel conditions~\cite{lin20215g, baeza2022overview}. These fluctuations often lead to variable levels of transmission signal loss. Under suboptimal conditions, the reception power can dip below -100~\!dBm ~\cite{hoyhtya2012application}, which in turn, severely compromises the spectrum sensing abilities of single satellite.


In comparison to spectrum sensing conducted by single satellite, the collaboration among multiple satellites enables simultaneous analysis of sensing data under diverse channel conditions. This significantly mitigates the impact of channel instability on the performance of spectrum sensing. Fig.~\ref{fig:teaser} illustrates a scenario of collaborative spectrum sensing (CSS) among LEO satellites. In this scenario, collaborative satellites capture signals from earth. Subsequently, each satellite individually transmits the sensing data to ground station (GS) for data fusion. Based on the results of this spectrum sensing, SUs are granted access to unoccupied spectrum.


However, the following limitations may be encountered when attempting to perform accurate and timely CSS: (1) Due to the varying degrees of Doppler shift experienced by each satellite‘s channel, the radio frequency (RF) channels between collaborative satellites exhibit heterogeneity. This heterogeneity leads to significant variations in the sensing data among the satellites, posing challenges for data fusion at GS. (2) Sensing a wide spectrum not only imposes significant sampling demands on analog-to-digital converter (ADC) but also generates a vast amount of electromagnetic data. On the other hand, due to the limited capacity of the satellite-to-ground link, transmitting such large volumes of data to the GS for data fusion in a timely manner becomes challenging.
(3) Data transmission from satellites to the GS often experiences packet loss due to extrem weather conditions and electromagnetic interference. Packet loss can impact the GS in conducting accurate and reliable data analysis, subsequently affecting the performance of spectrum sensing.

To address the above limitations, we propose a multi-satellite based CSS framework named \name. For Limitation (1), \name construct a graph based on the sensing data from each satellite to characterize correlations among RF channels of different satellites. Subsequently, we utilize graph learning to further explore the correlations among RF channels, thereby obtaining more effective representations during the training process. This approach helps mitigate the effects of heterogeneity in RF channels on the reliability of spectrum sensing.
For limitation (2), \name introduce a hybrid data compression method based on sub-Nyquist sampling and autoencoder (AE). In this approach, we initially employ sub-Nyquist sampling to mitigate the burden on the ADC and concurrently reduce the volume of sampling data. Subsequently, for the obtained sub-Nyquist data, we utilize the encoder of the AE to further compress it into an embedding. This embedding is then transmitted to the GS, where the decoder of the AE recovers the raw data (i.e., the sub-Nyquist sampled data of each satellite) from the embedding. For limitation (3), the concept of contrastive learning (CL) and AE are integrated in \name , aiming to compensate the loss of information by minimizing the distance between embedding with packet loss and ideal embedding in the new representation.
\begin{figure}[t] 
\centering
\begin{subfigure}{0.24\textwidth}
  \centering
  \includegraphics[width=4.31cm,height=3.87cm]{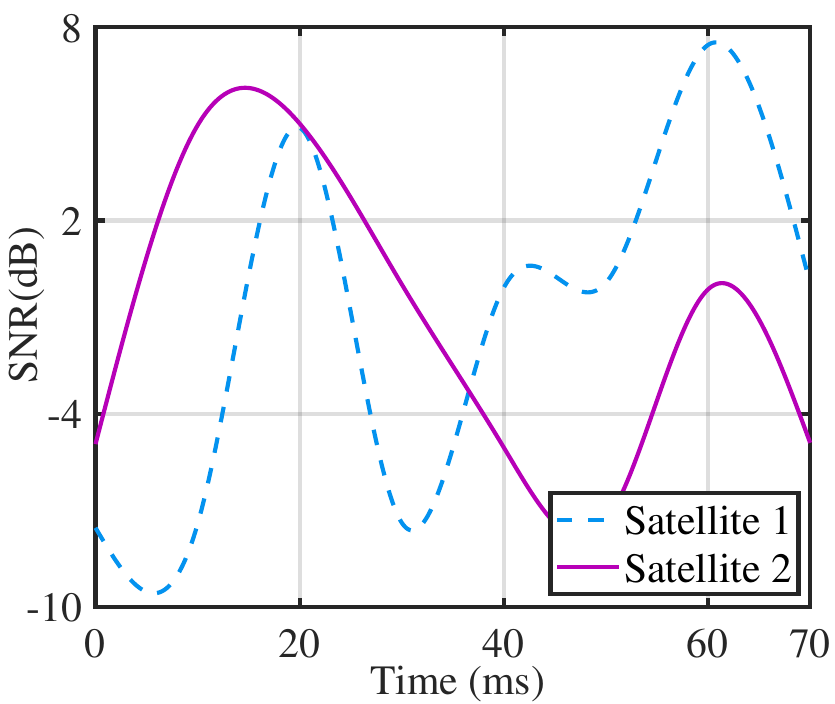}
  \caption{SNR}
\end{subfigure}%
\begin{subfigure}{0.24\textwidth}
  \centering
  \includegraphics[width=4.31cm,height=3.87cm]{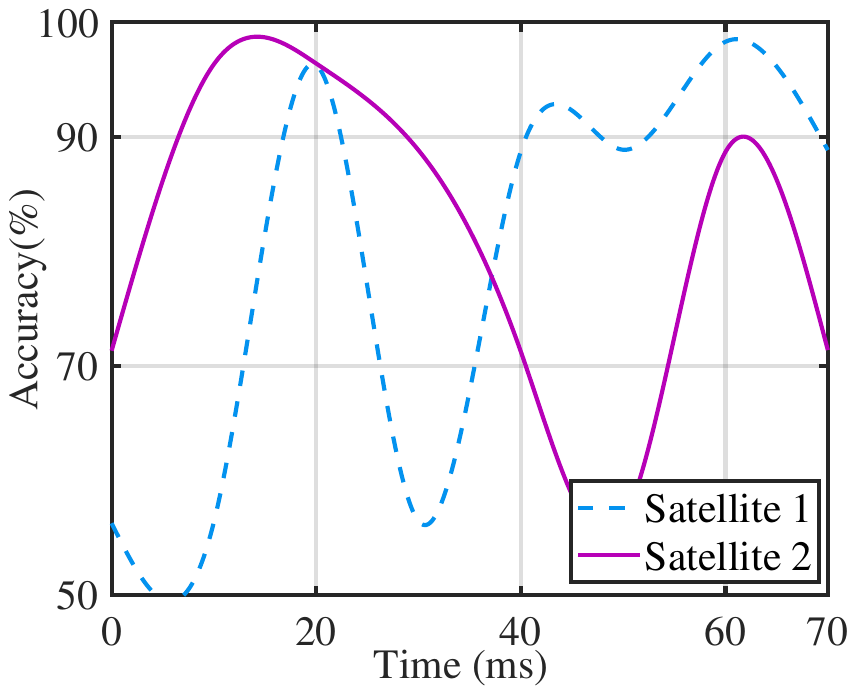}
  \caption{Spectrum sensing accuracy}
\end{subfigure}
\caption{The curve showing the variation over time of SNR and the corresponding spectrum sensing accuracy for two satellites.}
\label{fig:two_sat}
\end{figure}


Our main contributions can be summarized as follows:
\begin{itemize}
    \item To the best of our knowledge, The STRAS we proposed is the first spectrum sensing framework that jointly considers satellite data downlink and data fusion in GS.
    \item Given the heterogeneity of the RF channel, we model the sensing data from each satellite as a graph, and propose a method based on graph learning to achieve accurate and reliable spectrum sensing.
    \item In response to the constraints of the satellite-to-ground link capacity, we propose a hybrid data compression framework based on sub-Nyquist and AE , ensuring that sensing data from each satellite can be promptly transmitted to the GS for data fusion. 
    \item Addressing the packet loss phenomenon during satellite-to-ground data transmission, we incorporate CL into AE to compensate for missing packets, mitigating the impact of packet loss on the subsequent spectrum sensing performed in the GS.
\end{itemize}

This paper is organized as follows. Sec.~\ref{sec:motivation} motivates the design of by revealing the challenges faced by multi-satellite based CSS. Sec.~\ref{sec:design} presents the system design of \name. Sec.~\ref{sec:implementation} introduces system implementation and experiment setup, followed by performance evaluation in Sec.~\ref{sec:evaluation}. Related works and technical limitations are discussed in Sec.~\ref{sec:related_work}. Finally, conclusion and future remarks are presented in Sec.~\ref{sec:conclusion}.


\section{Motivation and Background} \label{sec:motivation}

In this section, we first discuss the limitations inherent to spectrum sensing by single satellite and emphasize the advantages and the imperative for adopting a multi-satellite collaborative approach. Following that, we explore the inherent challenges that arise when implementing these collaborative approach.
\subsection{The Issues of Spectrum Sensing in Single Satellite}
Existing methods often rely on single satellites to perform spectrum sensing, as cited in various works~\cite{zhang2019spectrum,benedetto2020cognitive,tian2023ed,jia2020intelligent}. However, as previously noted in Sec.~\ref{sec:intro}, the long distances and variable channel conditions inherent to satellite-to-ground links can compromise the stability of spectrum sensing performance when using just one satellite.



To better understand the aforementioned background, we conducted an experiment, which record the channel conditions faced by two independent satellites, as well as the corresponding spectrum sensing accuracy over time. The accuracy is derived from a single-point spectrum sensing method~\cite{uvaydov2021deepsense} and experimental results are shown in Fig.~\ref{fig:two_sat}. It can be observed that the channel conditions for both satellites fluctuate significantly over time, leading to inconsistent spectrum sensing performance. It is very common that a single satellite undergoes unfavorable channel conditions, characterized by a low signal-to-noise ratio (SNR), which in turn adversely impacts the accuracy of spectrum sensing. 

Fortunately, the growing number of LEO satellites offers a solution to these limitations through collaborative sensing. Specifically, as demonstrated in Fig.~\ref{fig:two_sat}, there may be instances when the channel conditions for satellite 1 are suboptimal, while those for satellite 2 are much better. In such cases, an effective collaborative approach among multiple satellites allows for the simultaneous analysis of sensing data under varied channel conditions, substantially reducing the impact of channel instability on spectrum sensing accuracy. As the LEO satellite network continues to expand, the scope for such collaborative efforts will only increase, further mitigating the effects of individual satellite channel degradation on the overall performance of spectrum sensing. However, these collaborative approaches also come with their own set of challenges, which we will explore in the subsequent subsections.

\begin{figure}
    \centering
    \includegraphics[width=0.7\linewidth]{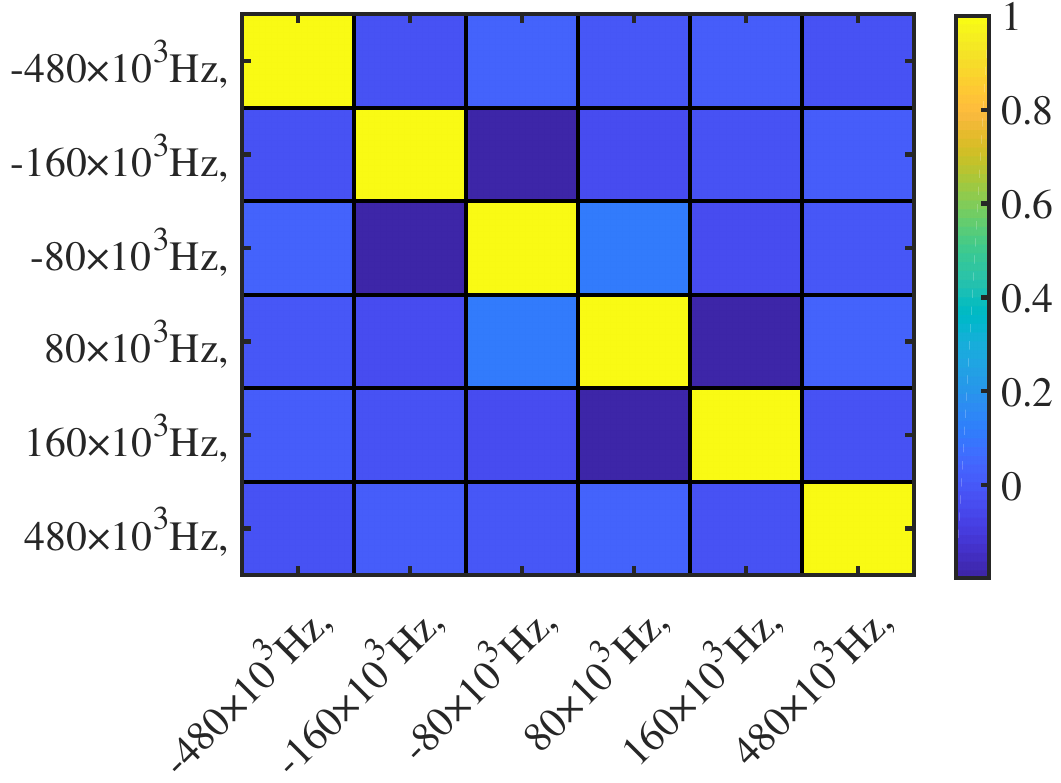}
    \caption{The Pearson coefficient of sensing data under different Doppler shifts.}
    \label{fig:pearson}
\end{figure}


\subsection{Heterogeneity between RF Channels}




When multiple satellites work in collaboration, they face the issue of heterogeneity in RF channels. This heterogeneity arises from several factors. First, at any given moment, collaborating satellites may be moving either towards or away from the user, resulting in considerable variations in Doppler shifts. Second, the angle between a satellite's trajectory and the direction of signal transmission to the user can also vary, directly influencing the Doppler shifts experienced. Lastly, satellites at different orbital altitudes travel at different speeds, further amplifying the inconsistencies in Doppler shifts.


This heterogeneity in RF channels is directly reflected in the sensing data collected by each satellite, leading to substantial discrepancies among these data sets. 
To empirically validate this observation, we calculate the Pearson coefficient between each sensing data at different Doppler shifts. The frequency of the transmitted signal ranges from 13.05~\!GHz to 13.8~\!GHz. According to the 3GPP protocol~\cite{3GPP2020}, the Doppler shift for each sensing data is set to range from -480 to 480~\!KHz. Fig.~\ref{fig:pearson} shows the results of the Pearson coefficient calculations. It is evident that the correlation between sensing data at different Doppler shifts is quite low. This low correlation further underscores that various Doppler shifts contribute to a significant discrepancies between sensing data.

Given the heterogeneity in RF channels, data fusion at the GS poses significant challenges. Specifically, when the sensing data from one satellite diverge markedly from those collected by other satellites, this discrepancy introduces a level of uncertainty into the aggregated data. Such uncertainty impairs the GS's ability to conduct accurate data analysis, thereby undermining the overall reliability of spectrum sensing. As a result, it is imprudent to treat all data from various satellites as equal during the data fusion process, simply concatenating them as inputs for CSS. Instead, an effective collaborative approach should be capable of considering the degree of mutual influence between the sensing data by establishing correlations between various sensing data, thereby, minimizing the adverse effects of heterogeneity of RF channels on spectrum sensing reliability.

\subsection{The Bottleneck in Satellite-to-ground Link Capacity}

Due to the statistical results of the L/S bands utilized by operational systems, dynamic spectrum sharing are compelled to sense a wide spectrum at wider frequency range to identify unoccupied spectrum~\cite{lagunas2015resource,li2017wideband}. However, sensing a wide range of spectrum poses two challenges. 

\begin{figure}[t]
  \centering
  \subfloat[A GS of Starlink.]{
    \includegraphics[height=3.3cm,width=3.8cm]{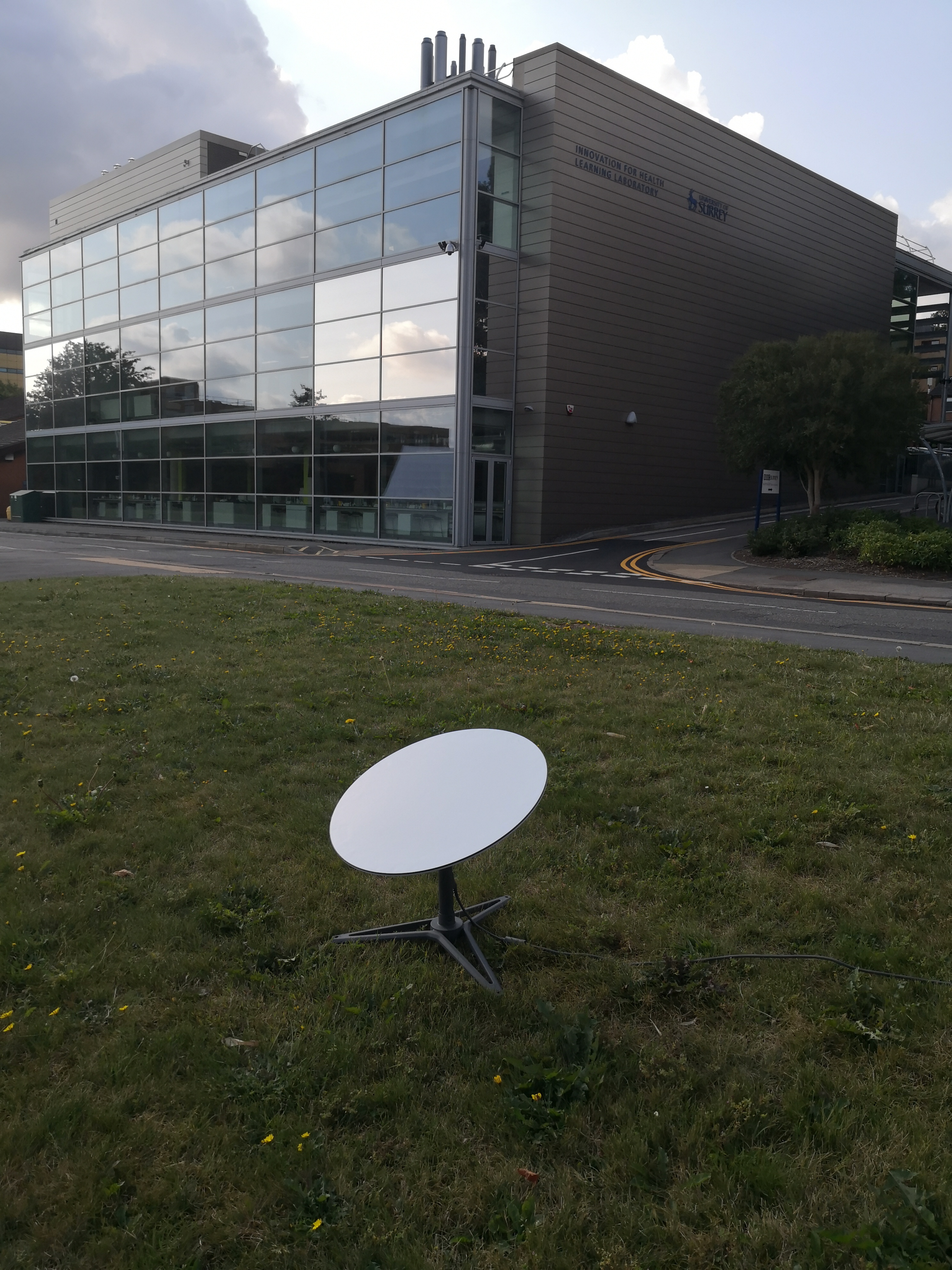}
    \label{fig:mtv:gs}
  }
  \subfloat[CDF of downlink rate.]{
    \includegraphics[height=3.3cm,width=4cm]{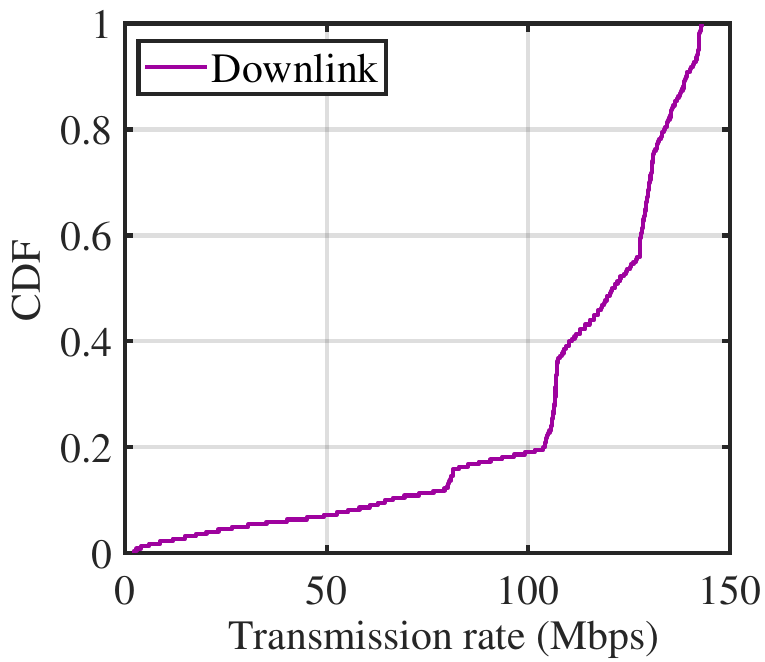}
    \label{fig:mtv:cdf}
  }
  \caption{The measured downlink transmission rate using the GS of Starlink.}
  \label{fig:motivation}
\end{figure}

Firstly, it demands an extremely high sampling rate on the satellite's ADC. Additionally, even if the ADC meets the sampling requirements, the sampling process will generate a huge amount of electromagnetic data that needs to be transmitted down to GS. As Table~\ref{tab:nyq_data} shows: the amount of IQ data generated from sampling a 1~\!GHz spectrum for 1 second is staggering. But the capacity of the satellite-to-ground link is inherently constrained, posing challenges for the timely transmission of sensed data from multiple satellites to the GS for CSS. We leverage an evaluation of the downlink rate utilizing the GS of Starlink in Guildford, England. Fig.~\ref{fig:mtv:gs} and Fig.~\ref{fig:mtv:cdf} respectively shows the GS of Starlink and the cumulative distribution functions (CDF) for the downlink rate. Our measurements indicate that the average downlink rate is approximately 100~\!Mbps. At this rate, it takes about 5 minutes to transfer electromagnetic data collected over a 1-second interval. This extended transmission time incurs significant time costs, posing a challenge for the prompt execution of data fusion.

\begin{table}[t]
\centering
\caption{The amount of data obtained from Nyquist sampling}
\begin{adjustbox}{width=0.5\textwidth}
\begin{tabular}{cccc}
\toprule
Sampling mode & Sampling rate & ADC precision & 1s sampling data volume \\
\midrule
Nyquist sampling & 2~\!GHz & 16~\!bit & 4~\!GB \\
\bottomrule
\end{tabular}
\end{adjustbox}
\label{tab:nyq_data}
\end{table}

\subsection{
    Packet Loss During Satellite-to-ground Data Transmission
} \label{sec:mtv:packet_loss}


In the course of data transmission from satellites to the GS, packet loss may occur~\cite{krishnan2004explicit,cianca2005integrated,chen2022satellite}. We also conduct a test on the packet loss rate using the GS of Starlink in Guildford, England. The average packet loss rate measured is 1.73\%. Moreover, due to the extended distance of the satellite-to-ground link and the high velocity of the orbiting satellites. the data transmission between the satellite and GS faces added challenges including, but not limited to, severe electromagnetic interference, unstable atmospheric conditions, and network congestion. Such complications heighten the risk of packet loss, undermining the reliability of the data received by the GS. This, in turn, affects the GS's ability to carry out dependable data analysis and subsequent spectrum sensing.

\section{Framework Design} \label{sec:design}

\subsection{Overview of \name Framework}

Motivated by Sec.~\ref{sec:motivation}, we propose a framework named \name to tackle the aforementioned limitations. The framework first constructs a graph based on the sensing data to characterize the correlations among RF channels of different satellites. Then, it employs graph machine learning algorithm to characterize the heterogeneity of RF channels across satellites, thus improving the effectiveness of model training. However, the limited downlink transmission rate poses a significant obstacle for GS to download massive electromagnetic data from satellites. To overcome this issue, we devise multi-coset sub-sampling and AE to reduce the sampled data volume and compress the sampled data. Furthermore, we integrate the CL into AE structure to compensate for the missing information caused by packet loss during the data transmission from satellite to GS. 

\begin{figure}[t]
    \centering
    \includegraphics[width=1\linewidth]{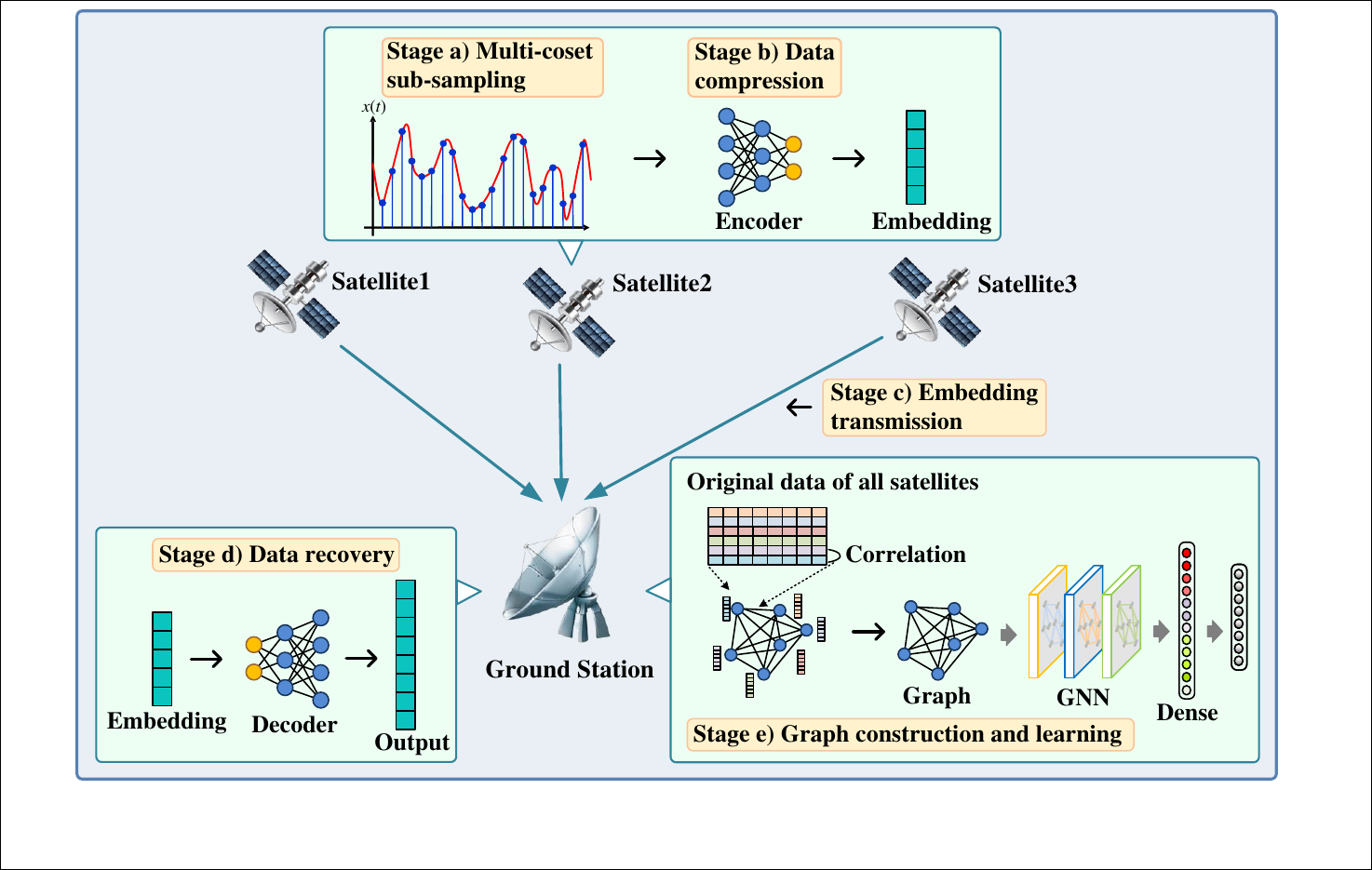}
    \caption{The overview of \name framework.}
    \label{fig:overview}
\end{figure}

\begin{figure*}
    \centering
    \includegraphics[width=0.9\linewidth]{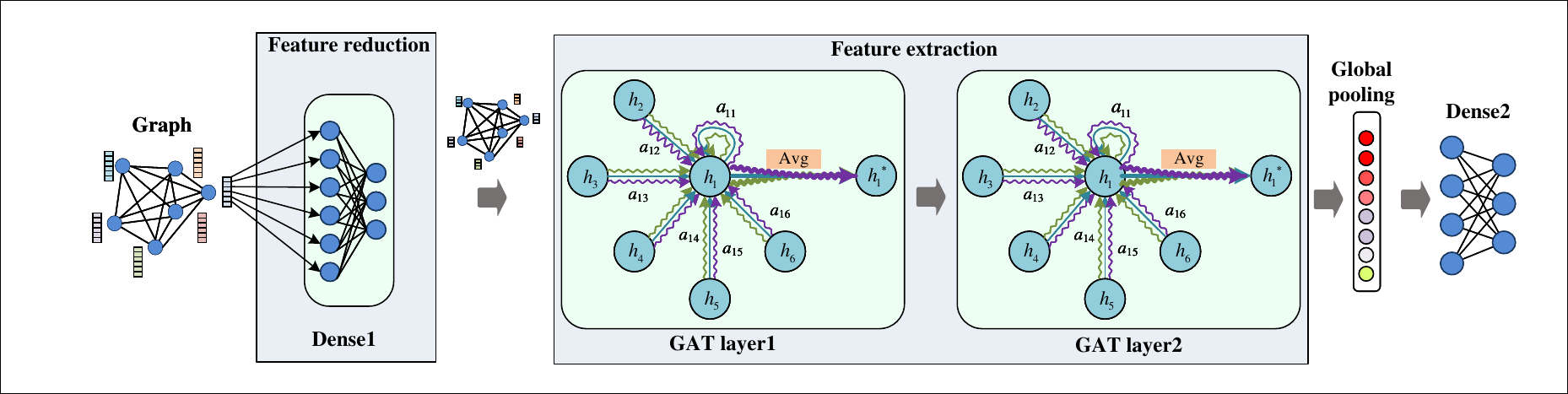}
    \caption{The architecture of GLSS.}
    \label{fig:architecture}
\end{figure*}

\subsection{Graph Learning based Collaborative Algorithm}

In the data fusion stage, we establish connections among various data by modeling the sensing data from each satellite into a graph. The weights of the edges reflect the degree of correlation between the sensing data.

For the modeled graph, GS can achieve reliable spectrum sensing by utilizing appropriate algorithms. However, due to the continuous movement of satellites, heterogeneity changes over time, characterizing the heterogeneity of RF channels within a unified  modeling scheme presents a formidable challenge. 

In recent years, with the rapid development of artificial intelligence, neural network (NN) have gained prominence as efficient feature extraction tool~\cite{sze2017efficient}. Through the powerful data feature extraction capabilities of NN, the heterogeneous information of satellite RF channels can be effectively extracted and analyzed. So, By utilizing graph neural network (GNN) on the constructed graph, the correlations can be further characterized, which helps to obtain the effective representation of each satellite, and achieve accurate spectrum sensing. The collaborative algorithm in \name can be divided into two steps: graph construction and graph learning.

\subsubsection{Graph Construction}

Sensing data from different satellites are modeled as a graph, denoted by $G(V, E, X)$. In this model, $V$ represents the nodes of the graph, comprising $K$ nodes $x_i\in V$, $X \in \mathbb{R}^{K \times2 P N}$ represents the node feature matrix. The feature dimension of each node is $2 P N$, we provide a detailed elaboration on the definitions of $P$ and $N$ in Section 3.3.1. $E$ signifies the edges between each node. Some existing works compute the received signal strength (RSS) of each frequency band as node features, and then used the Radial Basis Function (RBF) to calculate the correlation between nodes, subsequently assigning weights to the edges~\cite{zhang2022speckriging,janu2022graph}.


In our case, the entire framework relies on sub-Nyquist sampling samples, complicating the computation of RSS. Moreover, directly calculating the Pearson coefficient between the sensing data of each satellite as the edge weight becomes time-consuming when the sample length is excessive. Therefore, we model the graph as a fully connected, unweighted, undirected graph and let the graph learning model determine the edge weights. 

\subsubsection{Graph Learning}

For the constructed graph, we employ graph learning to explore the relationships among node features. GNN are classical graph learning methods based on deep learning~\cite{wu2020comprehensive,zhou2020graph}. Within the framework of a GNN, each node of the graph represents a sample, and each edge signifies the relationship between a pair of samples. The graph convolution operation learns the representation of graph nodes by propagating information between them, enabling the model to capture the complex relationships among the graph nodes. However, traditional GNNs lack the capability to compute edge weights between neighboring nodes; they necessitate that the graph already contains edge weight information.

Graph attention networks (GAT), by introducing the attention mechanism into graph learning, address the issue that traditional GNNs struggle to assign different weights to the edges of each neighboring node in the graph~\cite{velickovic2017graph}. Related to the GAT, we propose a graph learning based spectrum sensing (GLSS) algorithm, which consists of two FFNN (feed-forward neural network) layers and two GAT layers. The overall architecture of GLSS is shown in Fig. ~\ref{fig:architecture}.

To avoid overfitting due to excessively high dimension of node feature, GLSS initially uses FFNN to reduce the dimension of each node feature. As previously discussed, the input to GLSS is a fully connected graph with an adjacency matrix where all elements are set to one. Therefore, we also need to compute the weights of the edges to characterize the correlation between the nodes. For node $i$, the GAT layer first calculate the correlation coefficients between this node and its neighboring nodes $j \in \mathcal{N}_i$, as shown in the following formula:
\begin{equation}
  e_{i j}=a\left(\left[W^l h_{i}^l \| W^l h_{j}^l\right]\right), j \in \mathcal{N}_i
\end{equation}
where W is a shared weight matrix used to map the vertex features. $h_{i}^l$ represents the feature of the $ith$ node at the $lth$ layer. $[\cdot \| \cdot]$ represents the concatenation operation, where the results of mapping the features of two nodes are concatenated. 'a' is a learnable weight vector, which is constructed using a single-layer FFNN. The concatenated high-dimensional features are then mapped to a real number. After obtaining the correlation coefficients, normalize them using softmax to obtain attention coefficients, as shown in the following formula:
\begin{equation}
  \alpha_{i j} =\frac{\exp \left(\operatorname{LeakyReLU}\left(e_{i j}\right)\right)}{\sum_{k \in \mathcal{N}_i} \exp \left(\operatorname{LeakyReLU}\left(e_{i k}\right)\right)}
\end{equation}
The negative input slope of LeakyReLU is set to 0.2. After calculating the attention coefficients, we proceed with the feature aggregation operation. For each node, the new feature is obtained by taking the weighted average of its neighboring nodes' features. The weights for the weighted average are the attention coefficients calculated in the previous step. The entire process is illustrated as follows:
\begin{equation}
 h_i^{l+1}=\sigma\left(\alpha_{i, i} W^l h_i^l+\sum_{j \in \mathcal{N}_i} \alpha_{i, j} W^l h_j^l\right)
\end{equation}
where $\sigma$ denotes the Elu activation function. $h_{i}^{l+1}$ represents the feature of the $ith$ node at the ${l+1}th$ layer. Compared to calculating the attention coefficients of a node and its neighbors with one set of weights, GAT layer can use $T$ different sets of weights, where $T$ is the number of attention heads. For each set of weights, we calculate a set of attention coefficients and obtain a set of new node features. These features can be considered as different representations of information, each capturing different information within the graph. When calculating new node features, the features obtained from all attention heads are averaged, as shown in the following formula:
\begin{equation}
h_i^{l + 1} = \sigma \left( {\frac{1}{T}\sum\limits_{t = 1}^T {(\alpha _{i,i}^tW_t^lh_i^l + \sum\limits_{j \in {\kappa _i}} {\alpha _{i,j}^tW_t^lh_j^l)} } } \right)
\end{equation}
Where $\alpha _{i,j}^t$ represents the attention coefficients calculated using the $t$ $th$ set of weight $W_t^l$. After feature extraction through two layers of GAT, a global pooling operation is used to fuse the features of each node. Finally, the final output is obtained through a single layer FFNN, with the dimension of the output vector being equal to the number of sensing bands.
\subsection{Hybrid Data Compression}

\subsubsection{Multi-coset Sampling}
Compressed Sensing (CS) represents a signal acquisition and processing paradigm predicated on sub-Nyquist sampling rates. Through the utilization of CS algorithms, sparse signals can be effectively reconstructed from these sub-Nyquist samples~\cite{candes2008introduction,huang2021compressed,huang2022noncoherent}. The multi-coset sampler, a renowned compressed sampling architecture, has been extensively researched for the feasible and comprehensible implementation~\cite{mishali2010theory,moon2015wideband,ma2016reliable,yang2021adaptive,song2022approaching}. 


\begin{figure}
    \centering
    \includegraphics[width=0.8\linewidth]{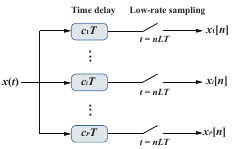}
    \caption{Multi-coset sub-sampling framework.}
    \label{fig:sub_sampling}
\end{figure}

To overcome the challenges posed by the large volume of downlink transmission data  and the limitation of existing ADC in sampling wideband signals due to its limited sampling frequency, we employ multi-coset sub-sampling approach. This approach utilizes multiple ADCs with unique time delays to perform parallel sampling of the signal. This allows sampling signals with  wider spectrum than a single ADC can handle, enabling  more accurate analysis and interpretation of complex signal data.

As illustrated in Fig.~\ref{fig:sub_sampling}, $x(t)$ is band-limited original signal within the range $\left[ {0, B} \right]$, where $B$ represents the maximum frequency of the band-limited signal. The Nyquist sampling interval is denoted by $T=\frac{1}{2B}$. In this multi-coset sub-sampling approach, the signals are parallelly sampled by $P$ ADCs and each ADC is configured with same sampling interval $LT$ ($P < L$),  where $L$ is a constant representing the ratio of the ADC sampling interval to the Nyquist sampling interval. Consequently, the sampled sequence of the $j$-th ADC is given by
\begin{equation}
{x_j}\left[ n \right] = x\left( {nLT + {c_j}T} \right),\quad n = 1,2,...,N,
\end{equation}
where $c_j$ denotes a non-negative integer satisfying $0 \le {c_j} < L$ and ${c_j} \ne {c_k}\left( {\forall k \ne j} \right)$. The sub-Nyquist samples obtained by multi-coset sub-sampling is denoted by $\mathbf{y} \in \mathbb{R}^{2 P \times N}$, where $N$ represents the number of sampled data for each ADC.

\subsubsection{Deep Autoencoder}

Despite multi-coset sampling decreases the amount of electromagnetic data transmitted, the data volume of these sub-Nyquist samples is still substantial in comparison to the downlink transmission rate. Therefore, to ensure timely transmission of the data to the GS for spectrum sensing, we need further data compression operation.


The essence of data compression is to identify and exploit patterns and redundancies within data. Wideband spectrum sampling generates large-scale, multi-type electromagnetic data, which is difficult to process efficiently with traditional methods. NN serving as an effective tool for data processing has following advantages:

\textbf{(1) Feature extraction capability:} NN utilize their non-linear modeling capability through activation functions, enabling them to capture complex data structures and relationships. Multi-layer NNs extract the most critical information from raw data by learning high-level features, representing this information with less data. \textbf{(2) Rapid inference:} NN leverage the parallel computing ability of their nodes, which, when combined with modern GPU or other specialized hardware, allows for efficient processing of large volumes of data. This parallelism significantly enhances computational speed, enabling complex NN models to quickly respond to and process real-time data.

Autoencoder (AE) is an unsupervised learning algorithm designed for dimensionality reduction and feature extraction\cite{bank2023autoencoders}. Comprising an encoder and a decoder—both often implemented via NNs. The encoder is responsible for  compressing input into a low-dimensional embedding containing the feature information of input. The decoder reconstructs a high-dimensional approximation of the original input from this embedding. In our scenario, the satellite leverages the encoder to compress the sensing data containing RF channel information and subsequently transmits the compressed data (i.e., embedding) to GS. The decoder at GS is then employed to recover the raw data.

We present the process of data compression and recovery by AE as follow. For data compression, the sub-Nyquist samples $\mathbf{x}$ with dimension $(2 P \times N)$ are flattened into a vector with dimension $2 P N$ as the input. The encoder then learns a compact representation from the input data, mapping the original high-dimensional input data to a low-dimensional latent representation, to obtain embedding $\mathbf{z}$. The encoding process is denoted by
\begin{equation}
\mathbf{z} = f(\mathbf{x};{{\bf{w}}_e})
\end{equation}
where $f(\cdot)$ represents the mapping function between the input data $\bf{x}$ and the embedding $\bf{z}$, and ${\bf{w}}_e$ is the weights of the encoder. The encoder of AE consists of two layers of FFNN with weight matrices $w^1_e \in \mathbb{R}^{K_1 \times (2P N)}$ and $w^2_e \in \mathbb{R}^{M \times K_1}$, where $M$ is the dimension of embedding $\bf{z}$. The mapping function of the encoder can be expressed as:
\begin{equation}
  f(\mathbf{x};{\bf{w}}_e)=a_e^2\left(w_e^2 a_e^1\left(w_e^1 \mathbf{x}+b_e^1\right)+b_e^2\right)
\end{equation}
where $a_e^1(\cdot)$ and $a_e^2(\cdot)$ denote the activation function for the first and second layers of the encoder, which are set as ReLU functions in the AE framework.

For data recovery, The GS takes the received embedding as input and produces the output $\mathbf{\hat{x}}$ by the decoder, the process of decoding is given by
\begin{equation}
\mathbf{\hat{x}} = g(\mathbf{z};{\bf{w}}_d)
\end{equation}
where $g(\cdot)$ represents the mapping function between the embedding $\bf{z}$ and the recovered data $\mathbf{\hat{x}}$. The decoder is deployed by a FFNN with single layer, and its weight matrix is  $w_d \in \mathbb{R}^{(2 P N) \times M}$. The mapping function of the decoder can be written
as:
\begin{equation}
  g(\mathbf{z};{\bf{w}})=a\left(w_d a\left(\bf{x}\right)+b_d\right)
\end{equation}

\subsection{Contrastive Learning for Packet Loss}

Ideally, embedding can comprehensively capture the fundamental features of raw data, facilitating the decoder's recovery of the raw data from embedding. However, as discussed in Sec.~\ref{sec:mtv:packet_loss}, data transmission from a satellite to the ground may experience packet loss. Especially, during extreme weather conditions and electromagnetic interference, the packet loss rate will further increase, resulting in significant loss of information in the embedding. For the decoder at GS, the information loss in embedding significantly impacts the recovery of raw data. Thus, it is imperative to devise strategies for compensating the absence of embedding information.


In light of the aforementioned limitation, how to compensate for the lost information in embedding is key to the decoder's accurate recovery of the raw data.
Contrastive learning (CL) is a type of unsupervised learning, and the core idea is to minimize the representational distance between positive samples and maximize it between negative samples by contrasting positive and negative sample pairs. By contrasting a vast number of sample pairs, the model can effectively extract better feature representations from the data~\cite{chen2020simple, he2020momentum,grill2020bootstrap}.


In CL, positive samples are closer in their representation space, indicating that they contain similar information in that space. Inspire by this, we propose the contrastive autoencoder (CAE), which incorporates the idea of CL into AE.
To compensate for the information loss in embedding caused by packet loss, the smaller distance between the embedding and the corrupted embedding in the representation space is better. Consequently, the information contained in samples with smaller representation space distances becomes more similar, thereby facilitating the decoder in precisely recovering the raw data from the corrupted embedding.

\begin{figure}
    \centering
    \includegraphics[width=1\linewidth]{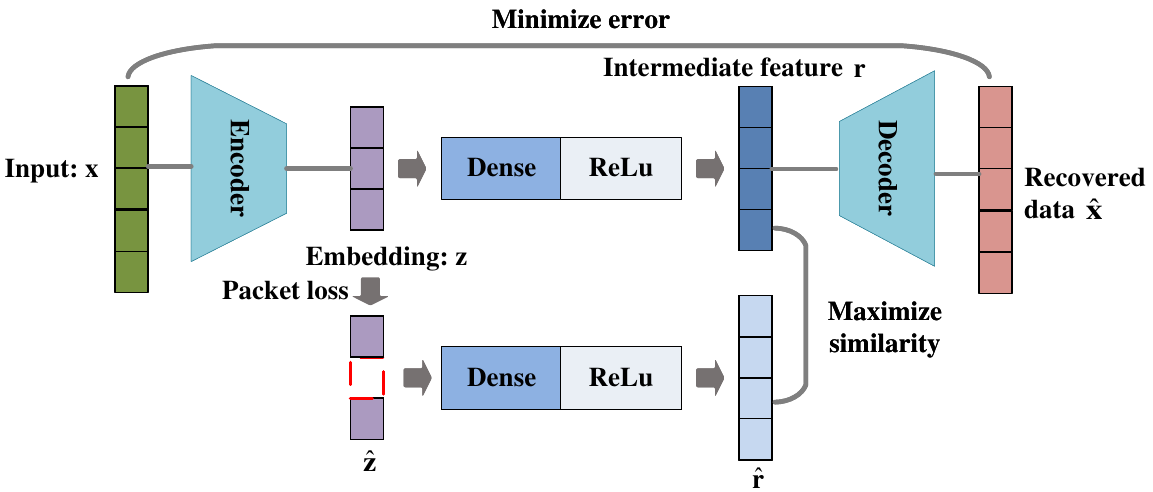}
    \caption{The architecture of CAE.}
    \label{fig:auto_encoder}
\end{figure}

As shown in Fig.~\ref{fig:auto_encoder}, after obtaining the embedding $\mathbf{z}$, we simulate packet loss in data transmission by discarding a portion of data from $\mathbf{z}$ based on the MPEG-TS encapsulation protocol~\cite{morello2006dvb}, resulting in a corrupted embedding $\mathbf{\hat z}$.  Subsequently, the intermediate features derived from $\mathbf{z}$ and $\mathbf{\hat z}$ through a single-layer FFNN are represented as
\begin{equation}
  \mathbf{r}=a\left(w_3 \mathbf{z}+b_3\right)
\end{equation}
\begin{equation}
  \mathbf{\hat r} = a({w_3}\mathbf{\hat z} + {b_3})
\end{equation}
where the $w_3 \in \mathbb{R}^{K_2 \times M}$, and $K_2$ represent the dimension of the weight. The final component of CAE is the decoder, the decoding process is denoted by
\begin{equation}
\mathbf{\hat{x}} = g(\mathbf{r};{\bf{w}})
\end{equation}

The loss function of the CAE utilizes the disparities between the recovered data $\mathbf{\hat x}$ and raw data $\mathbf{x}$, as well as intermediate features $\mathbf{r}$ and $\mathbf{\hat{r}}$,  to characterize the training loss. The first sub-loss function of CAE's loss function characterizes the difference between $\mathbf{x}$ and $\mathbf{\hat{x}}$ utilizing the Mean Squared Error (MSE), which is calculated as
\begin{equation}
  {L_1}\left( {\mathbf{x},\mathbf{\hat x}} \right) =\frac{1}{2 P N} \sum_{i=1}^{2 P N}\left\|x_i-\hat{x}_i\right\|^2
\end{equation}

The second sub-loss function of the CAE employs the cosine loss to measure the similarity between intermediate features $\mathbf{r}$ and $\mathbf{\hat{r}}$, which is expressed as
\begin{equation}
    {L_2}\left( {\mathbf{r},\mathbf{\hat r},k} \right)=
    \begin{cases}
    1-\cos \left(\mathbf{r}, \mathbf{\hat r}\right), & \text { if } k=1 \\
    \max \left(0, \cos \left(\mathbf{r}, \mathbf{\hat r}\right)-\operatorname{margin}\right), & \text { if } k=-1
    \end{cases}
\end{equation}
where $k$ is  an indicator variable that can take values of 1 or -1. If $k$ is 1, it indicates that input pairs are similar, and otherwise dissimilar. In our design, we aim to maximize the similarity between $\mathbf{r}$ and $\mathbf{\hat r}$, hence we set $k$ to 1.

Finally, we represent the loss function as the weighted average of the two sub-loss functions, which is given by

\begin{equation}
  {L}\left( {\mathbf{x},\mathbf{\hat x},\mathbf{r},\mathbf{\hat r},k} \right)=\alpha_1 {L_1}\left( {\mathbf{x},\mathbf{\hat x}} \right)+\alpha_2 {L_2}\left( {\mathbf{r},\mathbf{\hat r},k} \right)
\label{equ:cae_loss}
\end{equation}

\rev

\subsection{Putting All Together}

After discussing each function modules, we present the detailed workflow of \name consists of the following stages.

a) \textit{Multi-coset sub-sampling:} Satellites employ multiple ADCs to perform parallel sub-Nyquist sampling on the received signals. Furthermore, to enhance subsequent training stability, the sampled data is normalized using a Gaussian distribution. 

b) \textit{Data compression:} The normalized high-dimensional data is fed into the encoder of the CAE to obtain low-dimensional embedding, thereby achieving data compression.

c) \textit{Embedding transmission:} Satellites send the embedding to the GS. Given the potential occurrence of packet loss during data transmission, the embedding received by the GS may lose some information. 

d) \textit{Data recovery:} The GS inputs the corrupted  embedding into the decoder of the CAE for data recovery. 

e) \textit{Graph construction and learning:} The GS constructs a fully connected graph for the recovered data. The graph is then sent to the GLSS for spectrum sensing.

The pseudo-code for the \name framework is detailed in Alg.~\ref{workflow}.

\begin{algorithm}
    \caption{Workflow of \name framework}\label{workflow}
    \begin{algorithmic}[1]
        \STATE \textbf{Stage a):} Satellites conduct $p$ parallel sampling and normalize the sub-Nyquist samples to obtain $\mathbf{x} \in \mathbb{R}^{2 P \times N}$. 
        \STATE \textbf{Stage b):} Flatten $\mathbf{x}$ into a vector with dimension ${2 P N}$ as the input for the CAE encoder $f(\cdot)$, collecting the embedding $\mathbf{z} = f(\mathbf{x};{{\bf{w}}_e})$.
        \STATE \textbf{Stage c):} Satellite send the embedding $\mathbf{z}$ to the GS, where the received embedding is $\mathbf{\hat z}$.
        \STATE \textbf{Stage d):} GS use the embedding $\mathbf{\hat z}$ as input for the CAE and get the intermediate feature $\mathbf{r}$, then input it into the decoder $g(\cdot)$, and obtain the recovered data $\mathbf{\hat x}=g(\mathbf{\hat z})$.
        \STATE \textbf{Stage e):} The GS constructs a fully connected graph $G(V, E, \hat{X})$ for the recovered data. The graph $G$ is then sent to the GLSS for spectrum sensing.
    \end{algorithmic}
\end{algorithm}

\section{Implementation}
In this section, we elaborate on the dataset used in the experiment as well as the experimental settings for GLSS and CAE.
\label{sec:implementation}
\subsection{Dataset}

The whole sensing spectrum range from 13.025~\!GHz to 13.825~\!GHz and the possible carrier frequencies of transmitted singles range from 13.05~\!GHz to 13.8~\!GHz. Signals can be modulated using QPSK, 8PSK, or 16QAM. The number of singles is set to 2 and 3, with a bandwidth of 20~\!MHz each. In multi-coset sampling, the number of sampling channels $p$ is set to 8 with a sampling rate of 50~\!MSPS for each channel, the samples length for each channel $N$ is set to 400. As discussed in Sec.~\ref{sec:motivation}, we generate the sensing data of each satellites by adding different Doppler shifts and path losses according to the 3GPP protocol. The signals captured by satellites are also interfered with additive white Gaussian noise (AWGN). The number of collaborating satellites is set to 10, and the average SNR of all sensing data ranging from -10~\!dB to 10~\!dB.

\begin{figure}[ht]
\centering
\begin{subfigure}{0.25\textwidth}
  \centering
  \includegraphics[width=4.31cm,height=3.87cm]{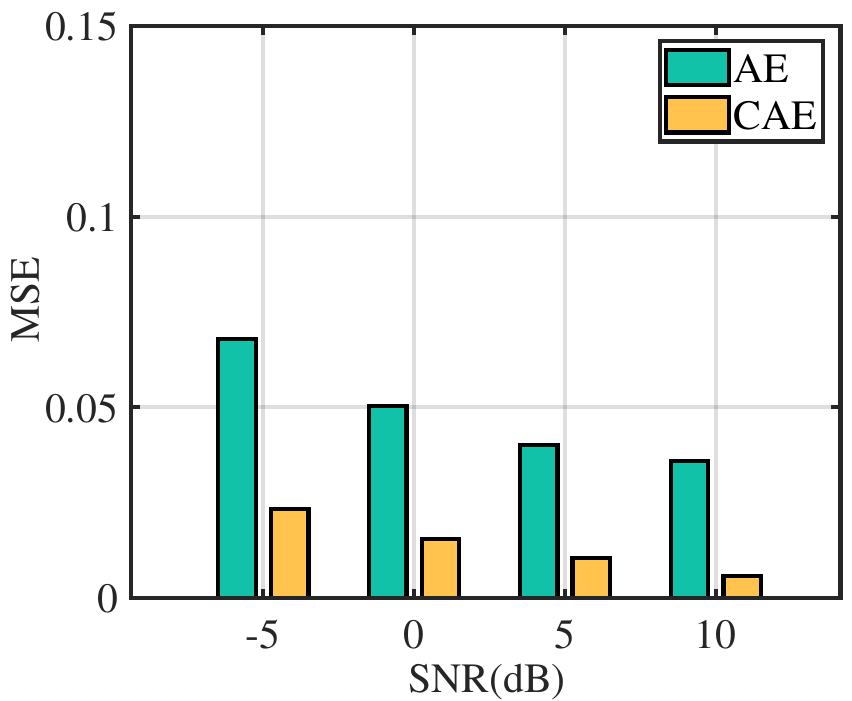}
  \caption{1\% packet loss rate.}
  \label{fig:sub:mse_1p}
\end{subfigure}%
\begin{subfigure}{0.25\textwidth}
  \centering
  \includegraphics[width=4.37cm,height=3.85cm]{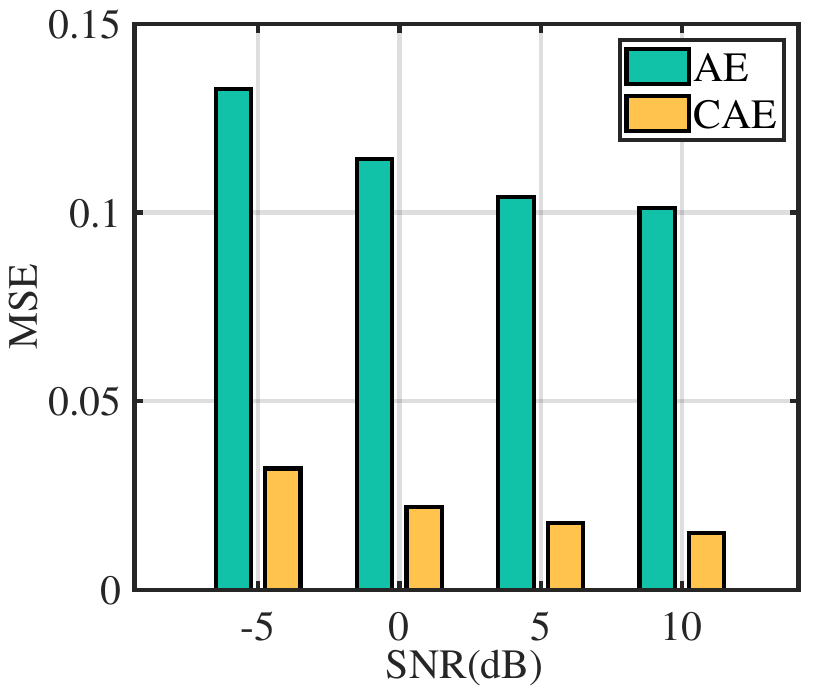}
  \caption{3\% packet loss rate.}
  \label{fig:sub:mse_3p}
\end{subfigure}
\caption{The comparison of MSE between CAE and AE at different packet loss rates and SNR.}
\label{fig:mse}
\end{figure}

\begin{figure}[ht]
\centering
\begin{subfigure}{0.25\textwidth}
  \centering
  \includegraphics[width=4.3cm,height=3.87cm]{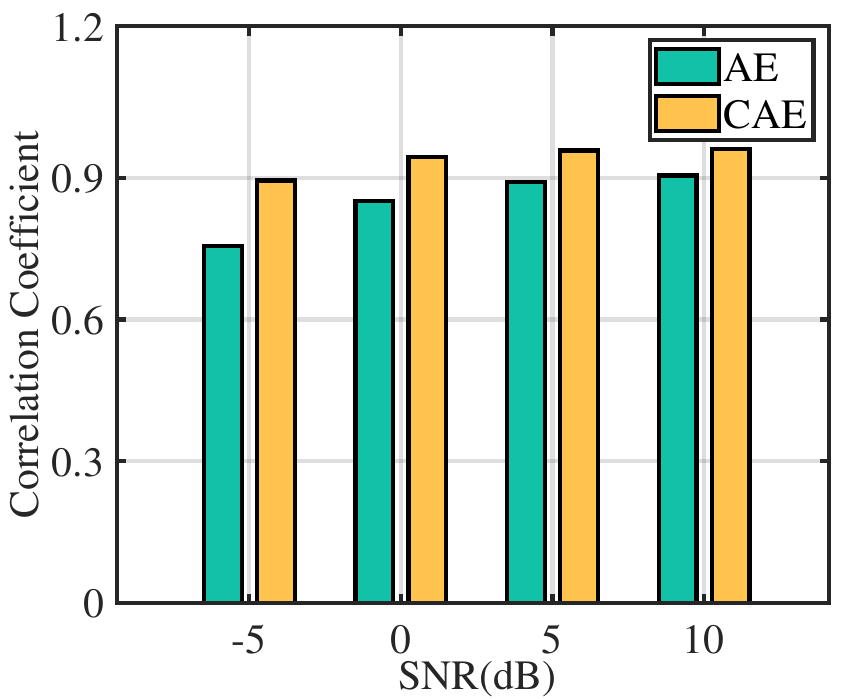}
  \caption{1\% packet loss rate.}
  \label{fig:sub:cc_1p}
\end{subfigure}%
\begin{subfigure}{0.25\textwidth}
  \centering
  \includegraphics[width=4.3cm,height=3.85cm]{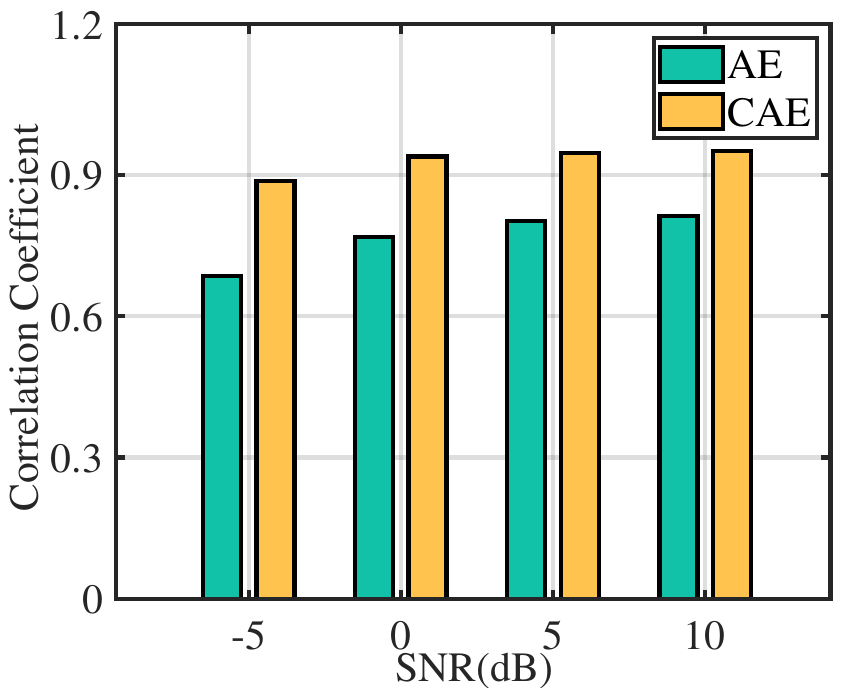}
  \caption{3\% packet loss rate.}
  \label{fig:sub:cc_3p}
\end{subfigure}
\caption{The comparison of the Correlation coefficient between CAE and AE at different packet loss rates and SNR.}
\label{fig:cc}
\end{figure}

\subsection{Graph Neural Network}

In the stage of spectrum sensing, we compared the performance of GLSS with a deep cooperative sensing (DCS) algorithm that utilizes convolutional neural network (CNN)~\cite{lee2019deep}. The loss function of GLSS and DCS is MSE and the training processes for spectrum sensing and raw data recovery both use the Adam optimizer~\cite{jais2019adam}, with an initial learning rate of 0.001. Additionally, to improve the performance by avoiding aggressive learning at the beginning when the model is rapidly changing, warmup~\cite{goyal2017accurate}, which increases the learning rate gradually at the beginning of 5 epochs is adopted and afterwards, the learning rate is adjusted using a cosine annealing schedule. The parameter configuration for the overall framework of GLSS is illustrated in Table~\ref{tab:GLSS_Parameter}.

\begin{table}[htbp]
  \centering
  \caption{Parameter of GLSS}
    \begin{adjustbox}{width=0.3\textwidth}
    \begin{tabular}{cccc}
    \toprule
    GLSS & \multicolumn{3}{c}{Layer} \\
    \midrule
    Dense1 & \multicolumn{3}{c}{FFNN(input\_dim, 640)} \\
    GAT1  & \multicolumn{3}{c}{GATconv(640, 256, heads=6)} \\
    GAT2  & \multicolumn{3}{c}{GATconv(256*6, 128, heads=6)} \\
    Dense2 & \multicolumn{3}{c}{FFNN(128*6, num\_classe)} \\
    \bottomrule
    \end{tabular}
    \end{adjustbox}
  \label{tab:GLSS_Parameter}
\end{table}


\subsection{Contrastive Autoencoder}

The accuracy of downstream spectrum sensing largely depends on the recovery Performance of the CAE on raw data, especially in the case of packet loss during satellite-to-ground data transmission. We first compared the performance of CAE and AE based on objective evaluation metric and subjective visual evaluation. The parameter configuration for the overall framework of CAE is illustrated in Table~\ref{tab:CAE_parameter}. AE consists of an encoder and decoder. The embedding dimensions of both AE and CAE are set to 640. The loss function of AE is MSE, while the loss function of CAE is given by Equ.~\eqref{equ:cae_loss}, where $\alpha_1$ and $\alpha_2$ are set to 1 and 3, respectively. 

\begin{table}[htbp]
\centering
\caption{Parameter of CAE}
\resizebox{0.3\textwidth}{!}{%
\begin{tabular}{ccccc}
\toprule
\multicolumn{2}{c}{CAE} & \multicolumn{3}{c}{Layer} \\
\midrule
\multicolumn{2}{c}{\multirow{2}[1]{*}{Encoder}} & \multicolumn{3}{c}{FFNN(input\_dim, 1600)} \\
\multicolumn{2}{c}{} & \multicolumn{3}{c}{FFNN(1600, embedding\_dim)} \\
\multicolumn{2}{c}{Dense} & \multicolumn{3}{c}{FFNN(embedding\_dim, 2048)} \\
\multicolumn{2}{c}{Decoder} & \multicolumn{3}{c}{FFNN(2048, input\_dim)} \\
\bottomrule
\end{tabular}%
}
\label{tab:CAE_parameter}
\end{table}

\section{Evaluation} \label{sec:evaluation}
In this section, we evaluate the performance of CAE and GLSS in \name, respectively, in raw data recovery and spectrum sensing.

\subsection{Performance for Raw Data Recovery}
In this section, we evaluate the performance of CAE based on two objective evaluation metrics: MSE and correlation coefficient, as well as subjective visual comparison. We also analyze the timeliness of CAE.
\subsubsection{Similarity Comparison between CAE and AE}
Fig.~\ref{fig:mse} depicts the MSE of the output and input data for CAE and AE across various packet loss rates and SNR conditions. In general, regardless of any SNR and packet loss rate conditions, the MSE of CAE is always significantly lower than that of AE. Even under the worst condition (-5 dB SNR, 3\% packet loss rate), CAE consistently outperforms AE operating under the best condition (10 dB SNR, 1\% packet loss rate). This demonstrates that with the help of CL, CAE is much more robust against noise. Also, CAE is less sensitive to packet loss. For instance, when the packet loss rate increases from 1\% to 3\%, the MSE of AE grows to more than twice its original value, whereas the MSE increase in CAE is relatively small. This suggests that under adverse network conditions (high packet loss rates), CAE's performance remains more consistent. The optimization of CL brings the mapping of embedding with packet loss closer to the mapping of original embedding, resulting its ability to resist the impact of packet loss. When we evaluate from another objective evaluation metric, it can be seen from Fig.~\ref{fig:cc} that CAE achieves a higher correlation coefficient than AE under any SNR and packet loss rate conditions. This further demonstrates that CAE is capable of producing outputs that are more similar to the raw data.



\begin{figure}[t]
  \centering
  \subfloat[1\% packet loss.]{
  \centering
\includegraphics[width=\linewidth]{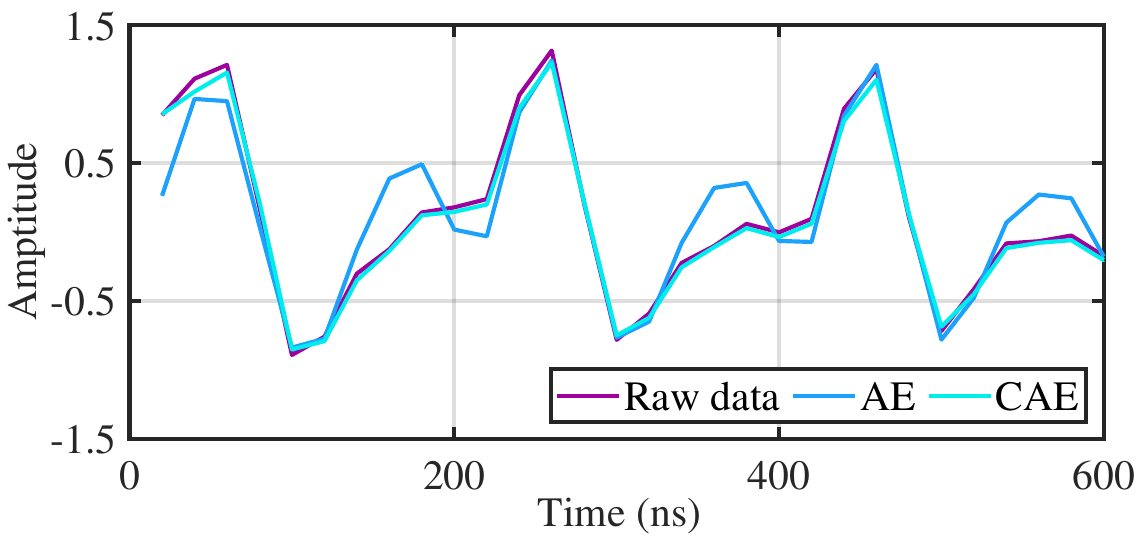}
    \label{fig:visual:1p}
  }\\
  \centering
  \subfloat[3\% packet loss.]{
  \centering
\includegraphics[width=\linewidth]{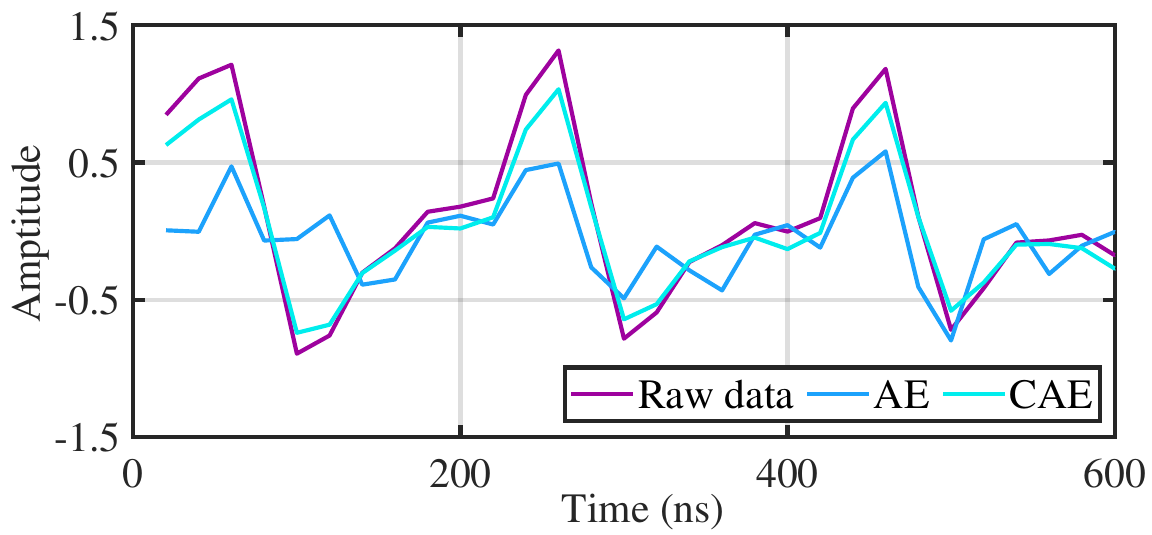}
    \label{fig:visual:3p}
  }
  \caption{Visual comparison between raw data and the outputs of CAE and AE at different packet loss rate.}
  \label{fig:visual}
  \vspace{-3ex}
\end{figure}
Fig.~\ref{fig:visual} shows a visual comparison of the outputs of CAE and AE with raw data under the conditions of 1\% and 3\% packet loss rate. As illustrated in Fig.~\ref{fig:visual:1p}, even the packet loss rate is only 1\%, after encoding and decoding, the output from AE is significantly distorted compared with the original input data. However, with the introduction of CL, CAE exhibits a high degree of fidelity, where the output is almost the same as the original input data. In Fig.~\ref{fig:visual:3p}, when the packet loss rate is 3\%, the distortion of AE is more severe in terms of both contour and amplitude. In contrast, the output of CAE still remains a considerable degree of similarity. From the visual perspective, CAE also demonstrates enhanced robustness, which makes it much better for more complex signal analysis tasks such as modulation classification.



\begin{figure*}[ht]
\centering
\begin{subfigure}{.33\textwidth}
  \centering
  \includegraphics[width=\linewidth]{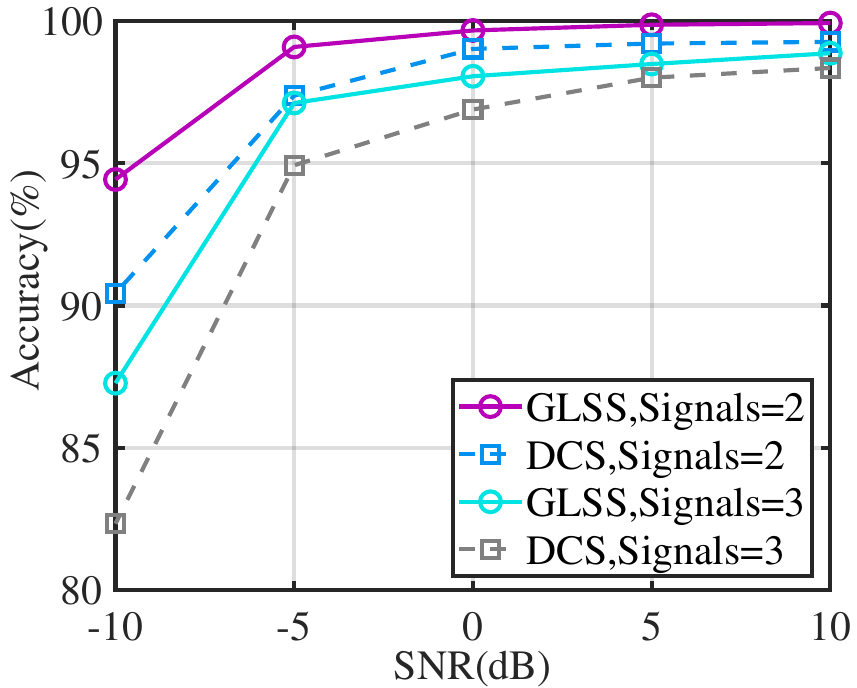}
  \caption{1\% packet loss rate.}
  \label{fig:acc_pkt_loss:p1}
\end{subfigure}%
\begin{subfigure}{.33\textwidth}
  \centering
  \includegraphics[width=\linewidth]{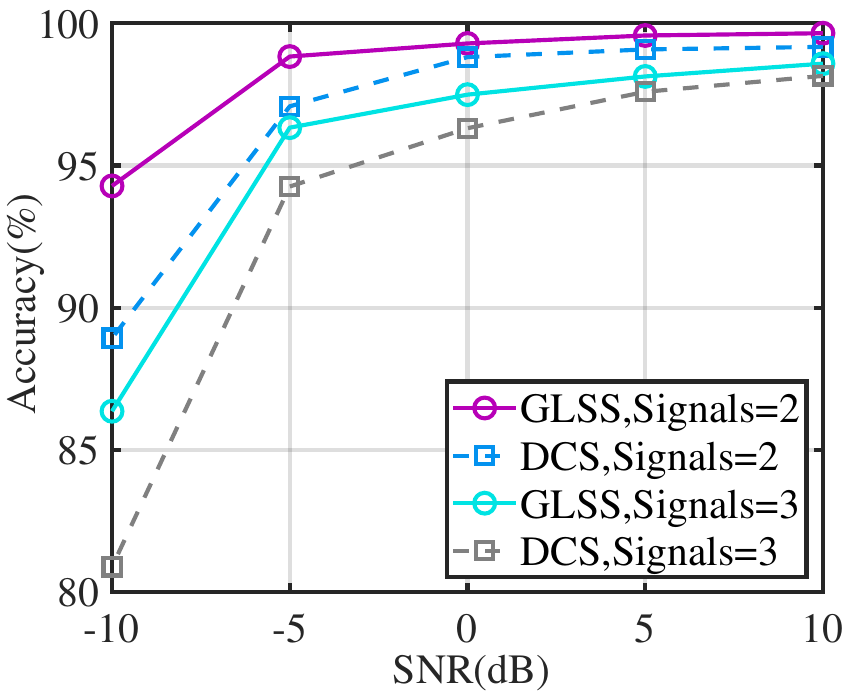}
  \caption{2\% packet loss rate.}
  \label{fig:acc_pkt_loss:p2}
\end{subfigure}
\begin{subfigure}{.33\textwidth}
  \centering
  \includegraphics[width=\linewidth]{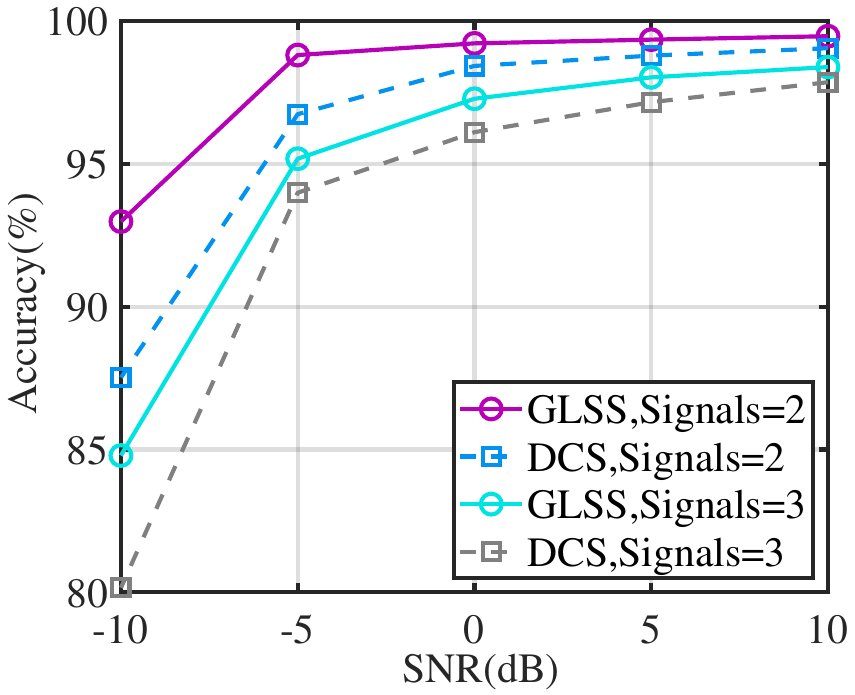}
  \caption{3\% packet loss rate.}
  \label{fig:acc_pkt_loss:p3}
\end{subfigure}
\caption{Accuracy comparison between GLSS and DCS at different packet loss rates and SNR.}
\label{fig:acc_pkt_loss}
\end{figure*}

\subsubsection{Timeliness Analysis}
The key to real-time spectrum sensing relies on whether the encoder of CAE can obtain the embedding from raw data and transmit it to the GS rapidly. Therefore, while evaluating the performance of the CAE, we also evaluate its timeliness. We use an NVIDIA RTX4090 GPU to process the data obtained from multi-coset sampling for 1 second. As shown in Table~\ref{tab:proc_time}, GPU memory and floating point operations per second (FLOPS) indicate the capabilities for handling large datasets and computational power respectively. With a batch size of 10,000, CAE is able to process the data sampled for 1 second in just 0.6576 seconds. This indicates that our CAE data compression scheme meets the real-time requirements. It can be anticipated that deploying CAE on specialized hardware will lead to even faster processing speeds.

\begin{table}[h]
\centering
\caption{Processing time of CAE}
\begin{adjustbox}{width=0.5\textwidth}
\begin{tabular}{cccc}
\toprule
GPU memory & FLOPs & batch size & Processing time \\
\midrule
24~\!GB & 82.58~\!T & 10,000 & 0.6576~\!s \\
\bottomrule
\end{tabular}
\end{adjustbox}
\label{tab:proc_time}
\end{table}

\begin{figure*}[ht]
  \begin{minipage}[b]{0.32\textwidth}
    \includegraphics[width=\textwidth]{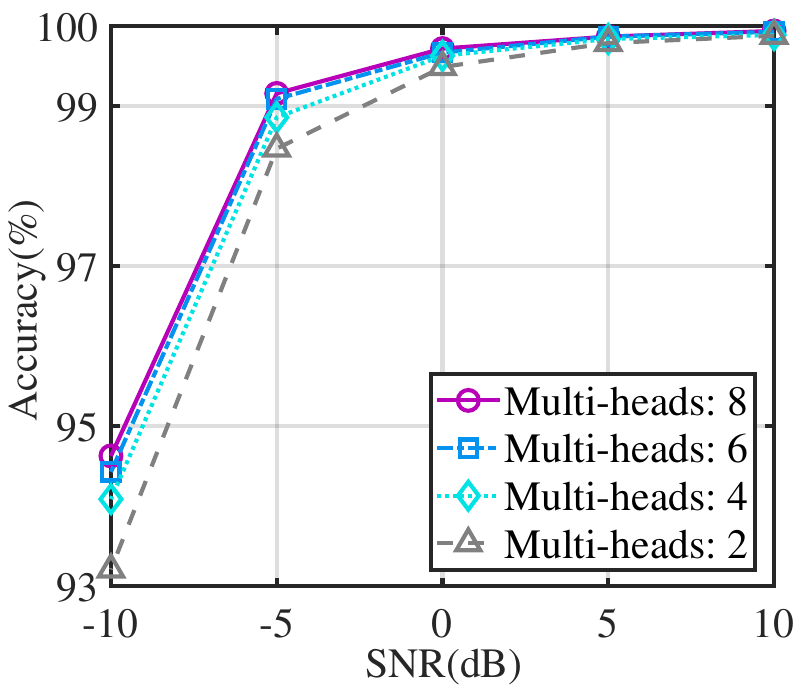}
    \caption{Impact of number of multi-heads on GLSS performance.}
    \label{fig:multi_head}
  \end{minipage}
  \hfill
  \begin{minipage}[b]{0.32\textwidth}
    \includegraphics[width=\textwidth]{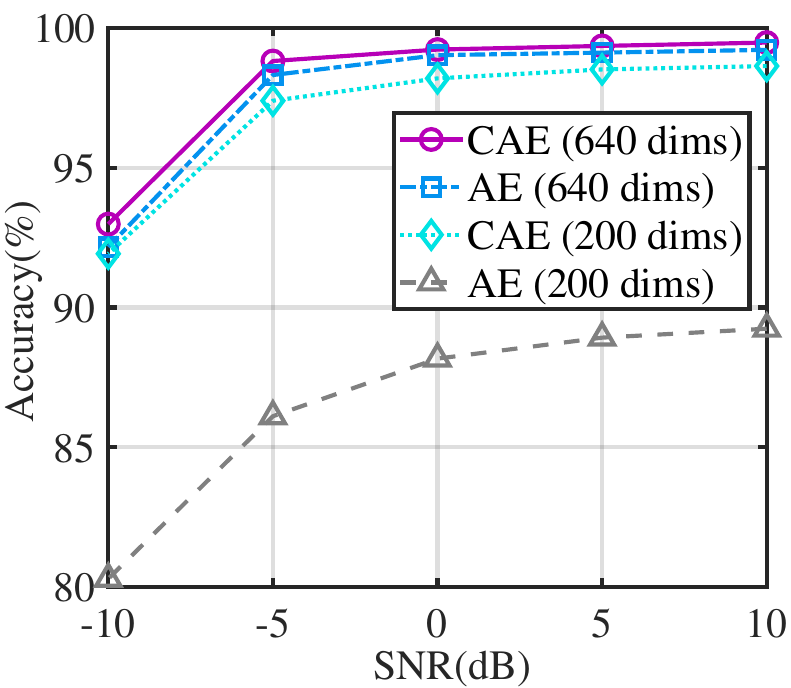}
    \caption{Impact of embedding dimensions on GLSS performance.}
    \label{fig:embed_dim}
  \end{minipage}
  \hfill
  \begin{minipage}[b]{0.32\textwidth}
    \includegraphics[width=\textwidth]{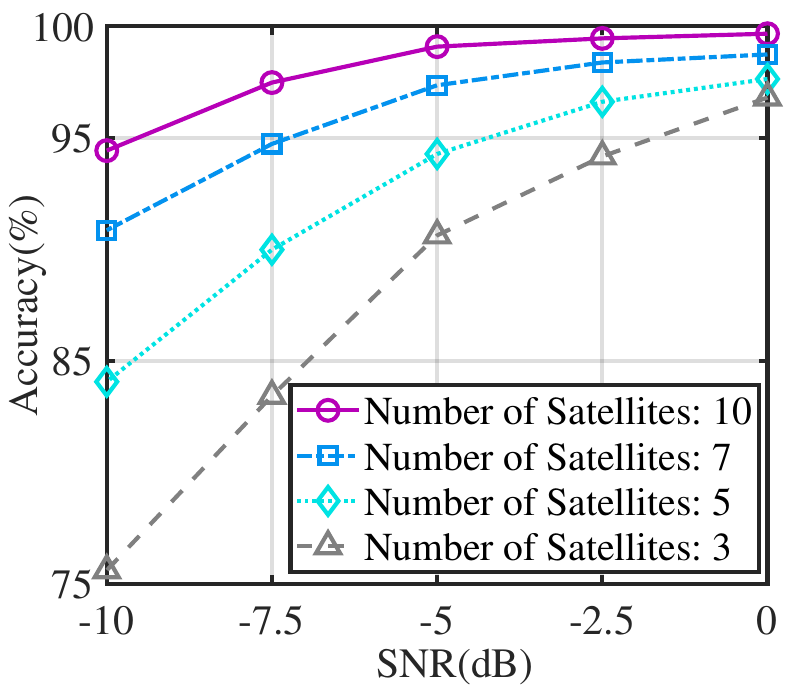}
    \caption{Impact of number of satellites on GLSS performance.}
    \label{fig:n_sats}
  \end{minipage}
\end{figure*}

\begin{figure}[t] 
\centering
\begin{subfigure}{0.24\textwidth}
  \centering
  \includegraphics[width=4.31cm,height=3.87cm]{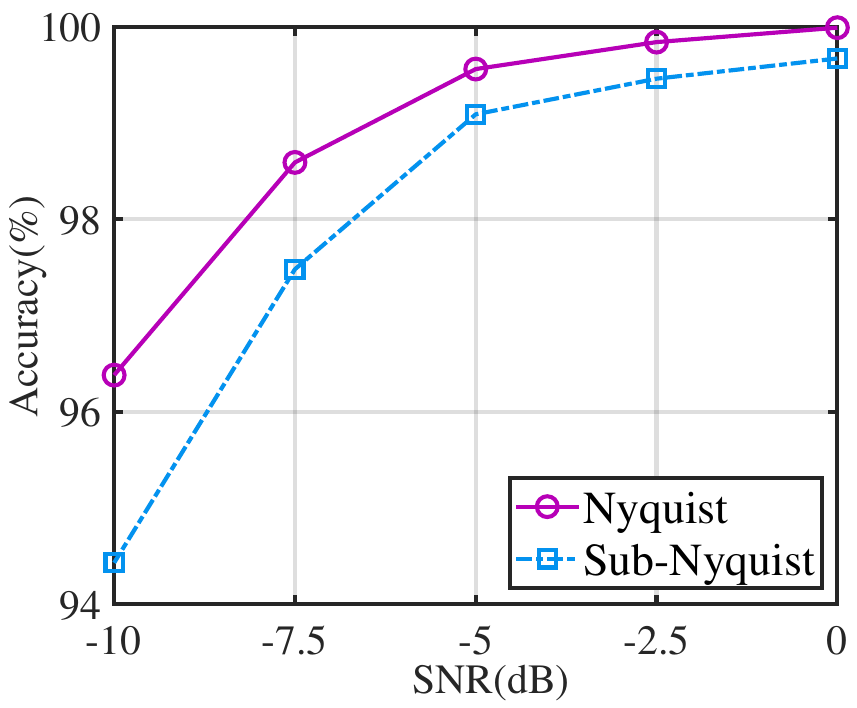}
  \caption{Comparison of Nyquist and sub-Nyquist sampling.}
  \label{fig:micro:nyq}
\end{subfigure}%
\begin{subfigure}{0.24\textwidth}
  \centering
  \includegraphics[width=4.31cm,height=3.87cm]{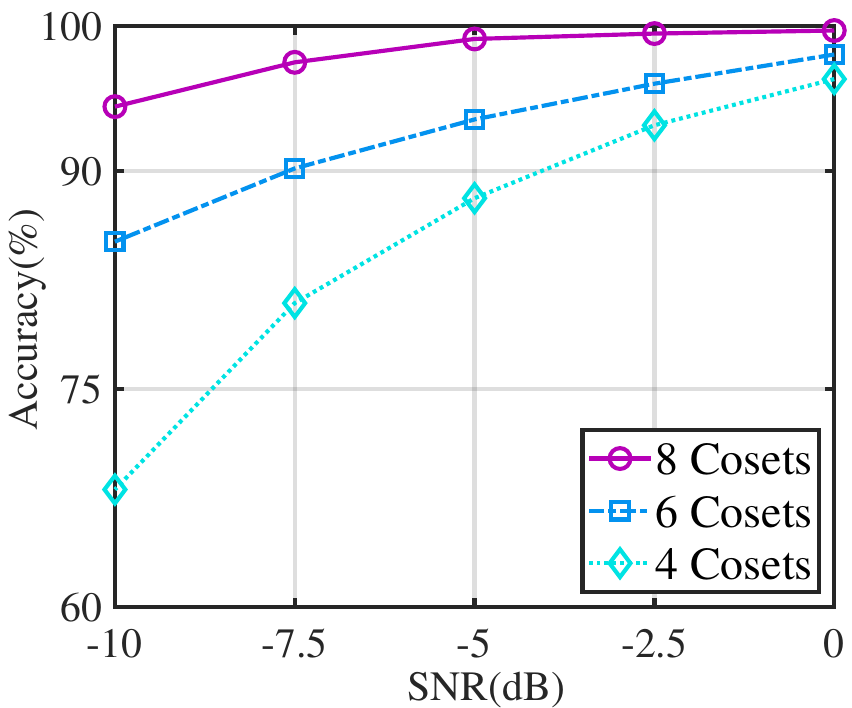}
  \caption{sub-Nyquist sampling with different numbers of cosets.}
  \label{fig:micro:coset}
\end{subfigure}
\caption{Impact of sampling mode on GLSS performance.}
\label{fig:micro}
\end{figure}


\subsection{Performance for Spectrum Sensing}
In this section, we first compare the spectrum sensing performance of GLSS to baselines. Subsequently, we evaluate the impact of different factors on GLSS performance.

\subsubsection{Accuracy Comparison between GLSS and DCS}

Fig.~\ref{fig:acc_pkt_loss} illustrates the accuracy of GLSS and DCS under
different SNRs and packet loss rates when the number of
signals in the spectrum is 2 and 3. In Fig.~\ref{fig:acc_pkt_loss}, the GLSS consistently outperforms the DCS model across all conditions of packet loss rates and SNRs. This superiority of the GLSS is particularly pronounced under low SNRs or high packet loss rates conditions. In particular, under -10dB SNR and 3\% packet loss rate, the accuracy of GLSS is about 5\% higher than that of DCS. Owing to the powerful feature extraction capabilities of neural networks, both GLSS and DCS exhibit better robustness at low SNRs. However, GLSS holds an advantage due to its GAT design, which enables it to capture and utilize inherent correlations within the sensing data for a more detailed analysis of channel heterogeneity. Consequently, this leads to more refined feature representations, thereby enhancing the fidelity of the spectrum sensing results produced by GLSS compared to DCS.

\subsubsection{Number of Multi-heads}

As previously discussed, multi-head attention can offer multiple ways to representation of data, thus enhancing the model's ability to extract the features of the input data and improve its performance.
We evaluate the influence of the number of multi-heads on the accuracy of GLSS. The packet loss rate is set to 1\% and there are 2 signals present in the spectrum. In Fig.~\ref{fig:multi_head}, as expected, GLSS achieves higher accuracy as the number of multi-heads increase. This can attribute to the fact that more multi-heads facilitate the model in capturing data correlations more efficiently. When the number of multi-heads is 2, the accuracy is relatively lower. When the number of multi-heads is above 4, the accuracy remains relatively consistent with no substantial performance improvement. We utilize 6 multi-heads in \name to strike a balance between accuracy and model complexity.



\subsubsection{Embedding Dimension}


The timeliness of GS spectrum sensing is influenced not only by the processing speed of the CAE discussed earlier but also by the embedding dimension. A lower embedding dimension means less data needs to be transmitted, thereby reducing the time for data transmission. However, a lower dimension makes it more challenging for the GS decoder to recover the original data, affecting the performance of spectrum sensing. To investigate the impact of the embedding dimension, we evaluate the performance of GLSS when the embedding dimension is reduced. The packet loss rate is set to 3\%, and results are shown in Fig.~\ref{fig:embed_dim}. When the embedding dimension is 640, this means the data volume is reduced by ten times compared to the original data dimension (2P×N) that needs to be transmitted. Although the output from AE exhibits considerable distortion, the decline in spectrum sensing accuracy is not particularly severe.

When the embedding dimension is reduced to 200, compared to the original data dimension, the compressed data volume is reduced by 32 times. At this point, the performance of AE significantly deteriorates. When the embedding dimension is low, AE fails to accurately recover the original data features from embeddings with information loss. In contrast, even at a low embedding dimension, CAE continues to exhibit relatively stable performance, further highlighting its robustness against packet loss. Compared to AE, due to the introduction of CL, CAE has greater capabilities for data compression, enhancing the timeliness of spectrum sensing in GS.
\begin{table}[h]
\caption{Computational complexity of GLSS} 
\centering
\begin{tabular}{ccc}
\toprule
Number of satellites & & FLOPs (million) \\
\midrule
10 & & 220.61 \\
7  & & 110.19 \\
5  & & 58.23 \\
3  & & 23.59 \\
\bottomrule
\end{tabular}
\label{tab:flops}
\end{table}

\subsubsection{Number of Satellites}

Fig.~\ref{fig:n_sats} represents the accuracy of GLSS for different numbers of collaborative satellites. By increasing the number of collaborative satellites, GLSS achieves improved spectrum sensing performance. Different from multi-head and the embedding dimension, reducing the number of collaborative satellites has a more pronounced impact on the performance of spectrum sensing. Following a reduction in the number of collaborative satellites from 10 to 5, the accuracy of spectrum sensing decreased by approximately 10\%. This indicates that, compared to single-point spectrum sensing, collaborative satellite spectrum sensing can significantly improve performance to address complex satellite communication scenarios. Therefore, spectrum sensing with extensive coverage area and high accuracy is made possible by leveraging collaborative satellites and GLSS.

However, the increase in the number of collaborative satellites also brings additional computational demands. In Table~\ref{tab:flops}, we present the floating point operations (FLOPs) of the models under different numbers of collaborative satellites. A trade-off needs to be considered between the accuracy of spectrum sensing and the computational cost of the model.


\subsubsection{Sampling Mode}

We also evaluate the impact of different sampling mode on GLSS. The results are shown in Fig.~\ref{fig:micro}. In Fig.~\ref{fig:micro:nyq}, it is evident that Nyquist sampling demonstrates superior overall accuracy compared to sub-Nyquist sampling. This is primarily due to the fact that sub-Nyquist sampling, despite alleviating ADC sampling pressure and reducing data volume, introduces data information loss, negatively impacting spectrum sensing accuracy. Extending our analysis, In Fig.~\ref{fig:micro:coset} shows that when the number of cosets is reduced to 6 or 4 from 8, there is a significant decrease in sensing accuracy, especially under low SNR conditions. We attribute this decline to the further loss of data information caused by the reduction in sampling channels, which impairs the model's ability to learn effective data representations and consequently degrades its robustness against noise.

\section{Related Work} \label{sec:related_work}
Deep learning (DL), utilizing NNs, can extract features from massive wireless communication data that are superior to manually constructed ones~\cite{lin2023efficient,chen2021rf,yuan2023graph,lin2023split,zheng2023autofed,qiu2024ifvit,lin2023pushing,fang2024automated,lin2024adaptsfl}. Based on the excellent data feature extraction capability of deep learning, the integration of spectrum sensing and DL has become even closer.
The authors in.~\cite{liu2019deep} have proposed a covariance matrix-based spectrum sensing algorithm, using the sample covariance matrix as input to a CNN, and demonstrated performance significantly superior to energy detection algorithms. 
Benefiting from the excellent feature extraction capabilities of DL, a real-time wideband spectrum sensing software/hardware framework have been proposed based on CNN~\cite{uvaydov2021deepsense}, which achieving precise spectrum sensing performance with a minimal amount of IQ data. Based on sub-Nyquist sampling , a wideband spectrum sensing method based on sub-Nyquist sampling has also been proposed by utilizing deep neural network (DNN) without the need for recovery algorithms~\cite{zhang2022machine}.

In addition to the aforementioned single-point spectrum sensing, DL also has extensive applications in CSS. The authors in~\cite{lee2019deep} have proposed a DCS framework based on CNN. In DCS, the sensing data of each node is concatenated to form integrated data. The CNN then extracts features from this integrated data to achieve high-precision spectrum sensing results. By combining DNN with an adversarial training database, ~\cite{liu2022attacking} effectively improved the ability to resist interference and attacks. The authors in~\cite{zhang2022speckriging} have used GNN to detect malicious nodes in collaboration, preventing false spectrum sensing information from affecting the overall performance of the CSS. 

With the rapid development of satellite communication, satellite spectrum sensing technology has garnered considerable attention in recent years. The authors in~\cite{benedetto2020cognitive} improve the performance of spectrum sensing by estimating the power of useful signals from the received noisy signals, using the second and fourth order moments of data. The authors in~\cite{zhang2019spectrum} have proposed a spectrum sensing and identification strategy using hypothesis testing and maximum a posterior probability to distinguish between GEO signals and interfering NGEO signals as well as noise. However, compared to CSS, the performance of the aforementioned spectrum sensing algorithm based on a single satellite is highly susceptible to channel fading and noise interference.

Compared to terrestrial networks, there is less research on CSS in satellite communication scenarios at present. The authors in~\cite{ding2022deep} combine CNN with long short-term memory (LSTM) to propose a spectrum sensing framework with low computational complexity. This framework has better noise robustness compared to the energy detection algorithm, but it is only designed for narrowband spectrum sensing tasks, and its performance in wideband scenarios has not been verified. The authors in~\cite{li2017wideband} have proposed a satellite wideband spectrum sensing technology based on CS, but this technology requires a reconstruction algorithm to obtain spectrum information, which increases the computational complexity of the framework and thus affects the real-time performance of spectrum sensing.

\section{Conclusion} \label{sec:conclusion}

In this paper, we propose \name, a spectrum sensing framework which jointly considers and addresses the challenges faced in multi-satellite based CSS, including channel heterogeneity, large-scale electromagnetic data downlink, and packet loss in satellite-to-ground data transmission. Such an approach enables fast and accurate spectrum sensing to become feasible. While the increase in the number of collaborative satellites enhances the detection accuracy of weak signals in spectrum sensing, it also leads to a rise in computational complexity. Therefore, striking a balance between computational complexity and detection accuracy requires more careful consideration. 




\ifCLASSOPTIONcompsoc
  \section*{Acknowledgments}
\else
  \section*{Acknowledgment}
\fi


\ifCLASSOPTIONcaptionsoff
  \newpage
\fi



%

\bibliographystyle{IEEEtran}
\bibliography{main}
\end{document}